\documentclass[conference]{IEEEtran}
\IEEEoverridecommandlockouts

\usepackage{cite}
\usepackage{multirow}
\usepackage{amsmath,amssymb,amsfonts}
\usepackage{algorithmic}
\usepackage{graphicx}
\usepackage{textcomp}
\usepackage{xcolor}
\usepackage{url}
\usepackage{array}
\usepackage{comment}
\usepackage{siunitx}

\newcommand{\gt}{\ensuremath >}
\def\BibTeX{{\rm B\kern-.05em{\sc i\kern-.025em b}\kern-.08em
    T\kern-.1667em\lower.7ex\hbox{E}\kern-.125emX}}

\usepackage{url}

\usepackage{breakurl}
\usepackage[breaklinks]{hyperref}

\begin{document}

\title{Design and Modeling of the Charge Readout of a SiMOS Quantum Dot with a Single Electron Transistor and CryoCMOS}

\author{Adam Quinn, Troy England, Andrew Li, Sharmila Mustari Nandita

\thanks{This work was produced by Fermi Forward Discovery Group, LLC under Contract No. 89243024CSC000002 with the U.S. Department of Energy, Office of Science, Office of High Energy Physics. Publisher acknowledges the U.S. Government license to provide public access under the DOE Public Access Plan DOE Public Access Plan.

A. Quinn is with the Fermi National Accelerator Laboratory, Pine \& Kirk St, Batavia, IL 60510
(e-mail: \protect\url{aquinn@fnal.gov}).

Report Number: PUB-26-0627-ETD}}

\maketitle

\begin{abstract}

Single electron spin qubits trapped in SiMOS quantum dots are a promising technology for scaling to thousand- or million-qubit systems due to their compatibility with mature CMOS manufacturing processes.  A readout system that combines a single electron transistor with a custom cryogenic CMOS amplification and digitization chain offers key advantages by avoiding the use of bulky RF components or room-temperature interconnects. We present design techniques and simulation results for an optimized qubit-SET cryoCMOS interface, culminating in the design of the QNDR1 ASIC, the first cryogenic readout ASIC designed under the Quandarum project, which targets the development of a many-channel spin qubit based detector for use in high energy physics. 


\end{abstract}

\section{Introduction}

\begin{figure}
  \centering
  \includegraphics[width=\linewidth]{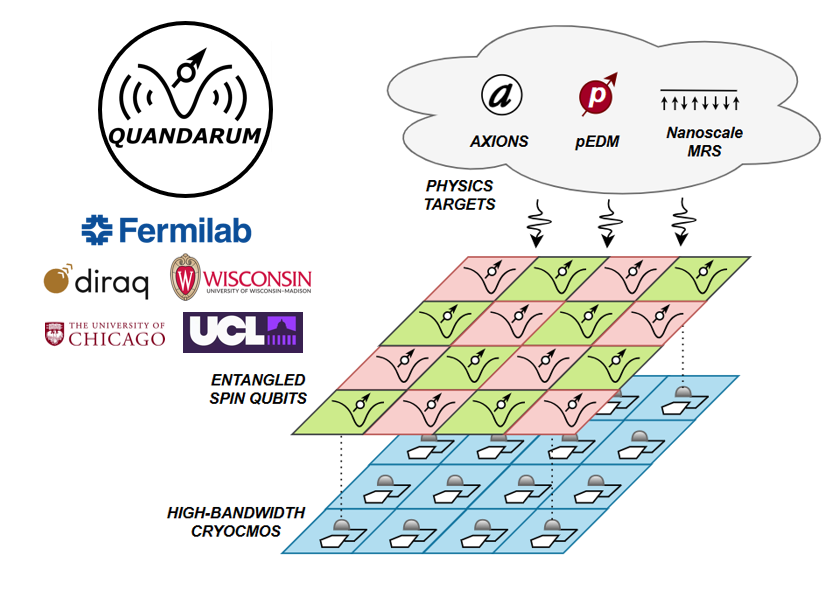}
  \caption{The objective of Quandarum is to advance the technology needed to deploy a large array of entangled silicon spin qubits with highly scaled cryogenic CMOS readout as a sensor to detect novel physics.}
  \label{fig:quandarum_graphic}
\end{figure}

In the search for physics beyond the Standard Model, High Energy Physicists are faced with a range of challenging questions about the nature of dark matter, symmetry violations, and the properties of fundamental particles, all of which demand the development of new experimental techniques. At the same time, progress in the field of quantum computing has made it increasingly practical to create intermediate-scale arrays of spin qubits, which can be used as exceptionally sensitive probes of quantum mechanical interactions \cite{Jackson_Kimball2023}. The goal of the Quandarum project is to combine spin qubits with physics expertise, creating a powerful and flexible technology for studying beyond standard-model (BSM) physics. (See Fig. \ref{fig:quandarum_graphic}.) 

In particular, electron spin qubits in silicon using CMOS technology offer significant promise in quantum sensing, primarily due to their compatibility with the existing semiconductor manufacturing infrastructure \cite{Steinacker2025}\cite{Bonen2018}. A combination of long intrinsic coherence times, a well-understood material environment, and small physical size means that silicon spin qubits have a clear pathway to scale to $10^4$ or more sensing elements. At this scale, quantum error correction (QEC) can be leveraged to maintain an ensemble of qubits in superposition for extended periods, allowing much more sensitive physics measurements through the spatio-temporal correlations of qubit disturbances.

To unlock the potential of spin qubits as sensing elements, a method for highly scalable, low-noise readout at cryogenic temperatures is essential. The state of the art readout technique for spin qubits in SiMOS quantum dots uses Pauli Spin Blockade (PSB) to convert the spin of a single qubit into a measurable charge displacement. A sensitive electrometer called a Single Electron Transistor (SET) is placed in the vicinity of the quantum dot and biased such that the charge displacement will cause a measurable change in the SET's conductance. The conductance of the SET is typically measured via RF reflectometry or by a transimpedance amplifier at room temperature. These approaches require the use of bulky RF components or room temperature interconnects which cannot scale to meet the readout needs of a practical fault tolerant quantum computer \cite{Guevel_2020}. 

A more scalable readout approach combines the single electron transistor with a custom cryogenic CMOS amplification and digitization chain. The cryoCMOS circuits can be integrated on a chip placed adjacent to or vertically integrated on top of the quantum dot circuit. Only a small number of room temperature interconnects are required for final readout of the digitized data, which can also be accomplished using optical fibers to further reduce heat load. However, implementing this readout approach requires careful co-design which takes into account noise, impedance matching, and cryogenic effects on all components.

In this paper, we present our approach to the design and modeling of the charge readout of a SiMOS quantum dot with a single electron transistor and cryoCMOS. The first two sections present our approach to modeling and optimizing the device-level performance of the single electron transistor and the cryoCMOS circuits respectively. Finally, we present the design of QNDR1, our cryogenic readout ASIC, along with simulation results. 

\begin{figure*}[!h]
  \centering
  \includegraphics[width=\linewidth]{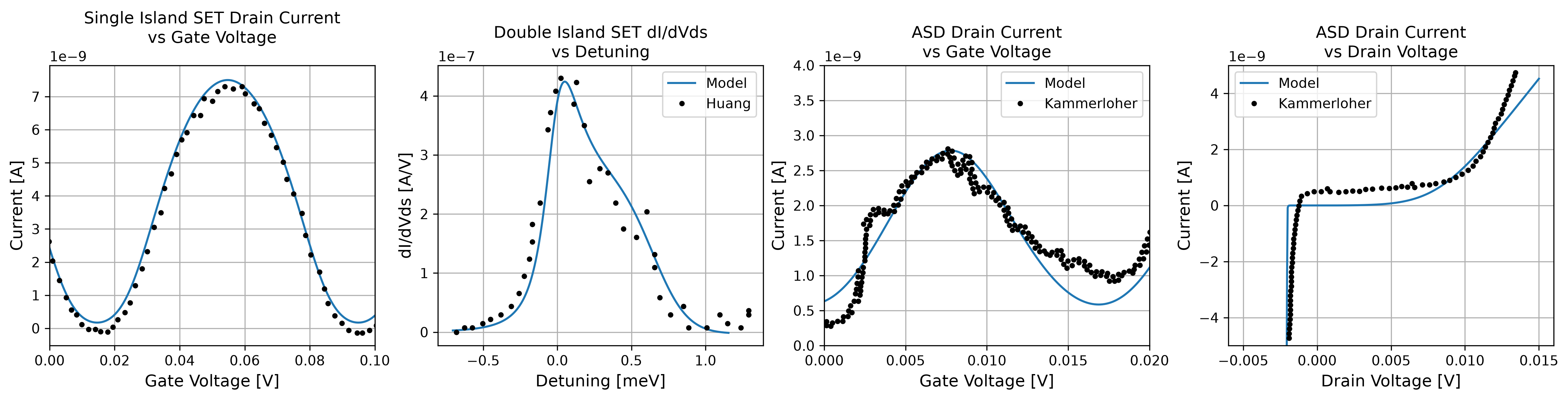}
  \caption{Plots illustrating the fitting of our Verilog-A models of the SISET, DISET, and ASD to data from the literature. In each plot, blue traces are simulated and black dots are data extracted from publications by Mahapatra \cite{Mahapatra2004}, Huang \cite{Huang}, or Kammerloher \cite{Kammerloher}.}
  \label{fig:set_fitting_plots}
\end{figure*}

\section{Single Electron Transistor Compact Modeling}
\label{sec:set_modeling}

Single Electron Transistors (SETs) are the technology of choice for reading out spin qubits via Pauli Spin Blockade due to their exquisite sensitivity. Elementary single-island SETs (SISETs) can be modeled effectively using the MIB (Mahapatra-Ionescu-Banerjee) model, a physics-based compact model \cite{Mahapatra2004}.

However, SETs suffer from many limitations including a large number of control voltages, sensitivity to noise and thermal broadening, and manufacturing variability. Several variations of the basic single electron transistor have been proposed to mitigate some of these limitations, but these variations lack canonical compact models like the MIB model, making it difficult to compare their advantages fairly or to co-design cryoCMOS readout of these devices.

We selected two promising alternative SET structures and developed Verilog-A models for them. We then co-simulated a SISET (using the MIB model) and our two alternative SET models with a basic CMOS readout circuit. The following subsections describe the two alternative SET structures and the results of our co-simulation. All of our Verilog-A models are available via the \textbf{qndr\_set\_modeling} repository on Github \cite{quinn_qndr_set_modeling}.

\subsection{Double Island Single Electron Transistor}

First, we selected the double-island single electron transistor (DISET), which consists of two quantum dots (or ``islands'') connected in series. Our model for the DISET is based on the work of Huang et al. \cite{Huang}, who demonstrated the DISET's potential as a high-sensitivity charge sensor that is relatively immune to thermal broadening in the source and drain leads. This advantage is due to the fact that the drain current of a correctly biased DISET depends primarily on the detuning between the left and right island potentials, rather than the potential difference between the islands and the source and drain leads. 

Huang gives a full mathematical treatment of the DISET transport characteristic based on the potentials of the source, drain, and left and right dots ($\mu_S, \mu_D,\mu_L,\mu_R$) but does not explicitly relate these potentials to terminal voltages. We developed a basic electrostatic model for the DISET based on lumped element capacitances analogous to the MIB model, which allowed us to fully derive the DISET current from terminal voltages. The parameter values for our model were derived from fitting to Huang's plot of $\frac{dI}{dV_{ds}}$ vs detuning as shown in Fig. \ref{fig:set_fitting_plots}.

\subsection{Asymmetric Sensing Dot}

Second, we selected the asymmetric sensing dot (ASD), a type of SET that incorporates an asymmetrically wide barrier between the quantum dot and the drain. This device was initially proposed by Kammerloher \cite{Kammerloher} with the objective of improving the output swing of SETs when used as charge sensors. The potential profile of the asymmetric drain barrier is controlled such that its tunneling resistance remains similar to the SISET, but the island-to-drain capacitance $C_D$ is dramatically reduced. In principle, this allows the ASD to have higher gain, easier biasing and greater robustness to kickback noise relative to a SISET. Previous experiments have reported an output signal as high as several millivolts from charge sensing experiments with an ASD, significantly higher than what is achievable with SISETs \cite{Kammerloher} \cite{Nielinger}.

The asymmetric drain barrier of Kammerloher's ASD is a compound structure consisting of a thin but tall tunnel barrier (Region I) and a long potential "slide" (Region II). Kammerloher models this barrier using finite element simulations but does not propose a compact model for the barrier or the ASD overall. 

To construct our model, we approximated Region I as a constant tunneling resistance $R_0$. Although the ``slide'' in Region II is not linear, we chose to approximate it as linear potential slope:

$$V(x) = U_0 + \frac{U_1-U_0}{W}x$$

Where $W$ is the total width of the slide and $U_0$ and $U_1$ are the potentials at the left and right endpoints. To best correspond with the data from Kammerloher, we treat $U_0$ as a fixed parameter and calculate $U_1$ as:

$$U_1 = \min(U_0, \mu_{d} +U_{1,offs})$$

\begin{figure}
  \centering
  \includegraphics[width=\linewidth]{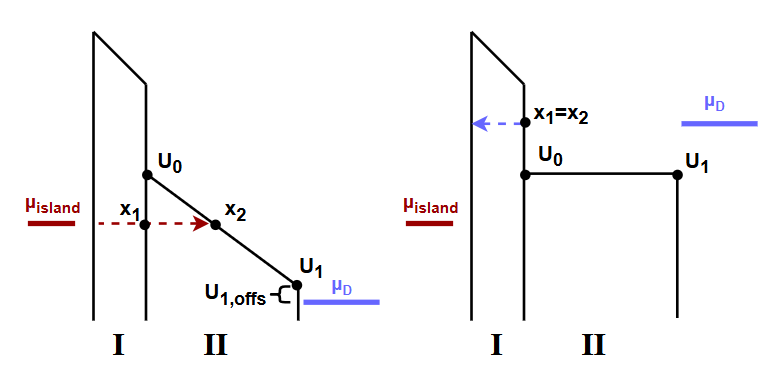}
  \caption{Potential diagram of the asymmetric drain barrier in our ASD model in two scenarios: (a) $\mu_D \ll \mu_{island}$ and (b) $\mu_D \gg \mu_{island}$.}
  \label{fig:ASD_potential}
\end{figure}

Fig. \ref{fig:ASD_potential} shows the potential of the asymmetric drain barrier in our simplified model.

Using the WKB approximation, the probability of tunneling through Region II alone is given by $T_{II} = e^{-2\gamma}$ where:

$$\gamma = \frac{1}{\hbar}\int_{x_1}^{x_2}\sqrt{2m(V(x)-E)}$$. 

In this expression, $m$ is the mass of an electron, $E = \max(\mu_{island},\mu_{D})$ is the maximum tunneling energy, and  ($x_1,x_2$) are the endpoints of the region where the electron is classically not allowed. Without loss of generality, we take $x_1=0$, and calculate $x_2$ as:

$$x_2 = W \frac{E-U_0}{U_1 -U_0}$$

With the constraint that:

$$0 \leq x_2 \leq W$$

We arrive at the closed-form solution:

$$\gamma = \frac{2W\sqrt{2m}}{3\hbar(U_1-U_0)} (U_0+\frac{U_1-U_0}{W}x_2 - E)^{3/2} - (U_0-E)^{3/2})$$

Using the Landauer formula, we can then find the overall tunnel resistance of the drain barrier, including both regions, to be:

$$R_D = \frac{R_0}{T_{II}}$$

Overall, our model for the tunneling resistance of the asymmetric drain barrier has four physical parameters ($U_0,U_{1,offs},W,R_0$). The rest of the ASD can be modeled using the MIB model with four additional parameters ($R_S,C_S,C_D,C_G$), yielding a novel eight parameter model for the transport characteristics of an ASD. We found the parameter values for our model by fitting to $I_D$ vs $V_G$ and $I_D$ vs $V_{DS}$ data presented by Kammerloher, as shown in Fig. \ref{fig:set_fitting_plots}.

The fit achieved by this model in Fig. \ref{fig:set_fitting_plots} is imperfect, which we believe is likely due to the simplifying assumptions we made, in particular regarding the linear shape of the Region II barrier. Future work is required to study if relaxing these assumptions can yield a higher quality compact model.

\subsection{SET-CMOS Co-Simulation}

We implemented the MIB model along with our new DISET and ASD models in Verilog-A. To our knowledge, this is the first time that it has been possible to simulate a circuit containing all three of these structures along with traditional CMOS transistors. 

\begin{figure}
  \centering
  \includegraphics[width=\linewidth]{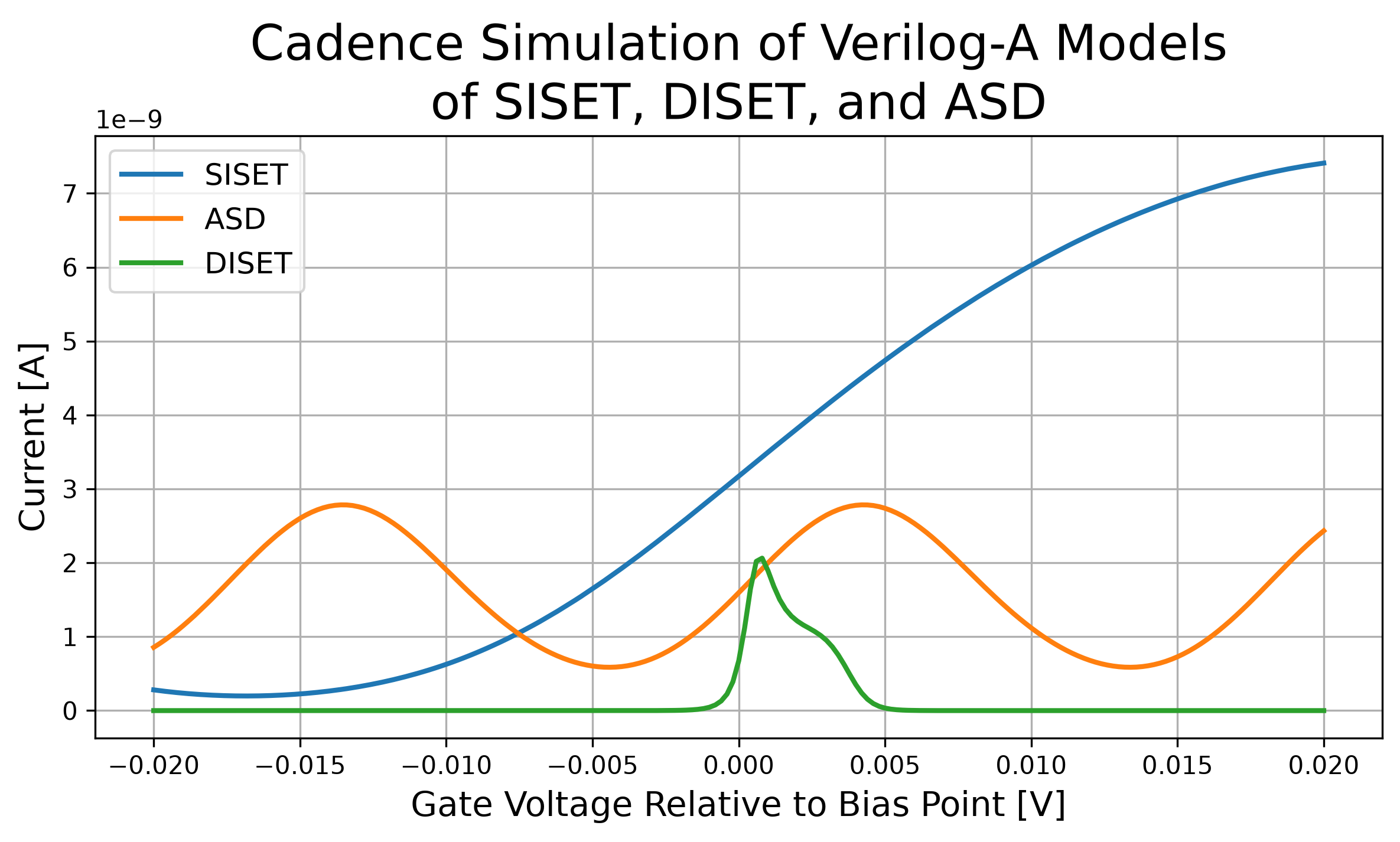}
  \caption{Coulomb peaks measured from simulating our Verilog-A models of the SISET, DISET, and ASD. $V_G=0$ represents the gate bias of the SET, which is selected to bias each SET flavor at maximum $dI/dV$.}
  \label{fig:set_iv_sweeps}
\end{figure}

The use of these models makes it simple to compare the performance characteristics of different SET structures, such as impedance and Coulomb peak height and steepness. Fig. \ref{fig:set_iv_sweeps} shows the simulated Coulomb peaks for all three SET models.

\begin{figure}
  \centering
  \includegraphics[width=0.8\linewidth]{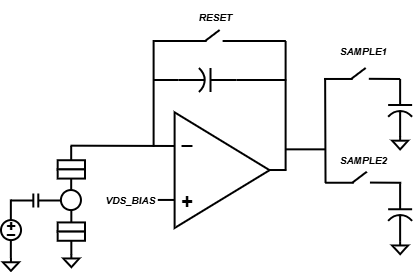}
  \caption{Simplified readout circuit used to evaluate our Verilog-A models of the SISET, DISET, and ASD.}
  \label{fig:set_test_schematic}
\end{figure}

\begin{figure}
  \centering
  \includegraphics[width=\linewidth]{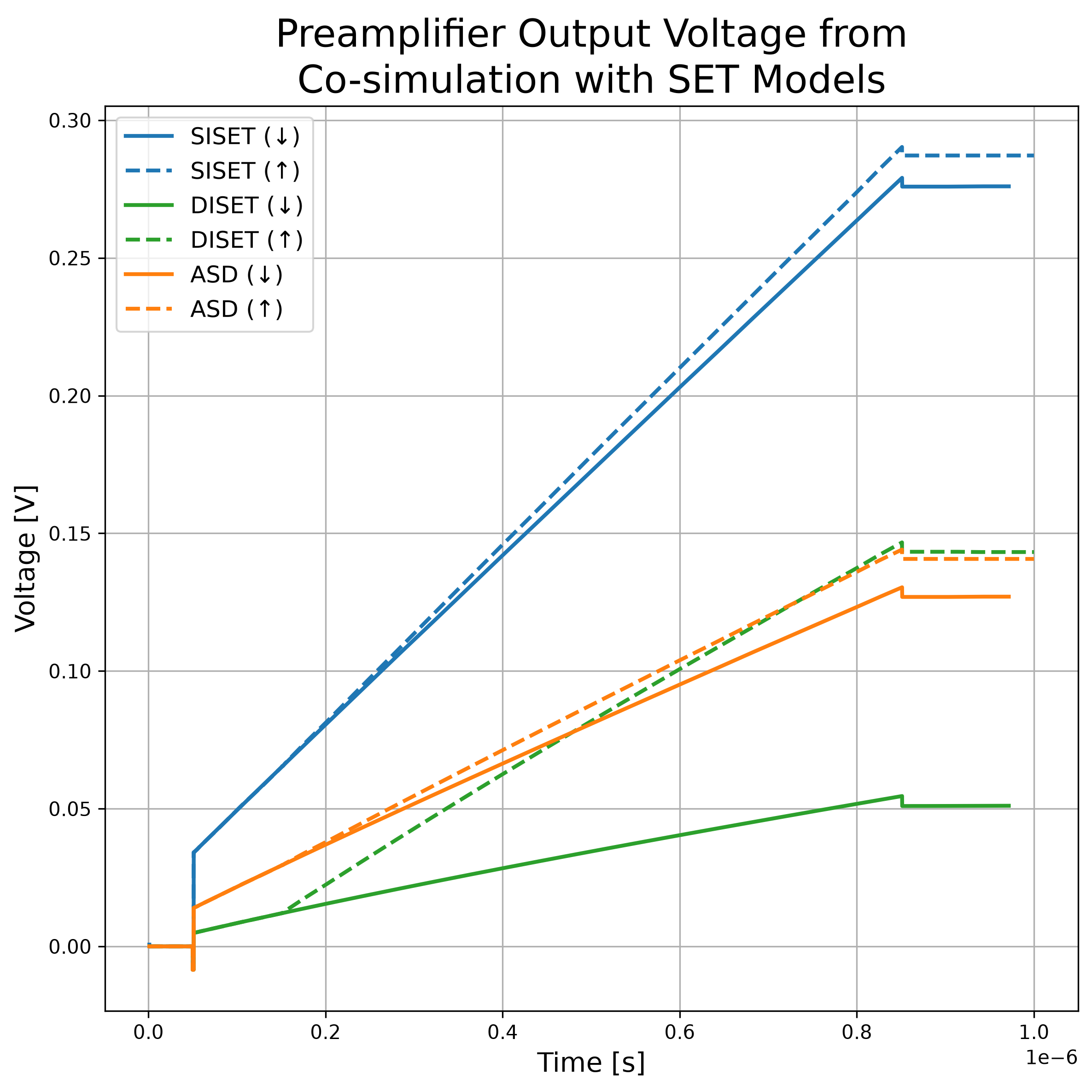}
  \caption{Cadence simulation of the readout of our Verilog-A models of the SISET, DISET, and ASD using the readout circuit in Fig. \ref{fig:set_test_schematic}}
  \label{fig:SET_cosim_sampled}
\end{figure}

It is also straightforward to co-simulate these models with CMOS readout. We connected each of the three SETs to a simple readout circuit consisting of a charge-sensitive amplifier and a correlated dual sampling circuit, as shown in Fig. \ref{fig:set_test_schematic}. We applied a $500 \unit{\micro\volt}$ signal to the gate of each SET and the integrated output signal from the preamplifier is shown in Fig. \ref{fig:SET_cosim_sampled}. In this simulation, the DISET produces a much larger output signal than the other two SET varieties due to the steeper Coulomb peak slope which is evident in Fig. \ref{fig:set_iv_sweeps}. However, it is important to note that the three specific devices here are based on three completely different literature sources (Mahapatra, Huang, and Kammerloher). The devices have different bias conditions and are not necessarily comparable.

For future work on Quandarum, these models will be fit to proprietary data from SET structures developed at Diraq. 

\begin{figure}
  \centering
  \includegraphics[width=0.8\linewidth]{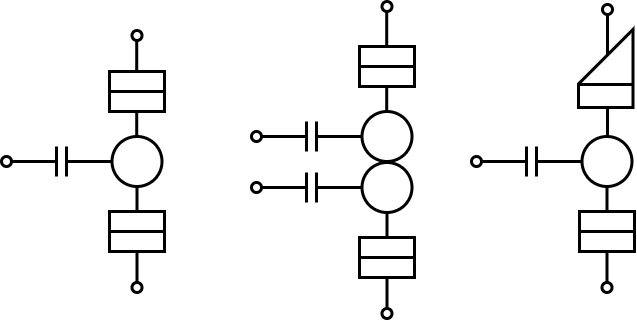}
  \caption{Proposed symbols for (a) a symmetric single-island SET (SISET), (b) a double-island SET (DISET), and (c) an Asymmetric Sensing Dot (ASD)}
  \label{fig:set_symbols}
\end{figure}

In addition to compact models, the DISET and ASD lack commonly recognized symbols. We propose the symbols shown in Fig. \ref{fig:set_symbols} 

\section{Cryogenic CMOS Design and Modeling}

A central challenge in the design of cryogenic integrated circuits is the lack of simulation models which take into account the shifts observed in CMOS device performance parameters at deep cryogenic temperatures, including threshold voltage and free carrier mobility \cite{Han_2021}. 

To partially address these challenges, we have developed an isothermal PDK for 22nm FDSOI operating at 3.8 Kelvin based on in-house measurements. This cryoPDK, presented previously in Ref \cite{Seidel2023}, includes extracted parameters for Super-Low Threshold (SLVT), Regular Threshold (RVT), and Enhanced-Gate Low Threshold (EGLVT) devices, with fitting errors of less than five percent for $I_{on}$ and $I_{off}$. As a consequence, the core circuits of QNDR1 are designed exclusively with SLVT, RVT, and EGLVT devices. 

However, some factors are not accounted for by this cryoPDK, including resistor superconductivity and noise analysis, which are addressed separately in the following sections. 

\subsection{Resistor Superconductivity}

Aside from transistor performance shifts, resistor superconductivity is an important challenge in the specific 22nm process used for this chip design. In general, unsilicided polysilicon resistors are known to maintain resistance close to their nominal value at several Kelvin \cite{Marques_Garcia_2024}. However, in this process, polysilicon resistors are manufactured with a thin TiN layer beneath them which may become superconducting when operated below a critical bias current \cite{quantum_machines}. Due to this possibility, the feedback resistor in the Version B preamplifier (discussed below) is implemented as a MOS resistor with a floating voltage bias determined by a small 2-bit current DAC. Elsewhere in our design, polysilicon resistors are used for applications such as source degeneration and level shifting, where a modest bias current is assumed.

\subsection{Cryogenic Noise Performance}
\label{sec:cryo_noise}

There are two difficulties in simulating the noise performance of QNDR1 at cryogenic temperatures. First, we do not have accurate noise models for devices operating at cryogenic temperature, as noise measurements were outside the scope of the Fermilab cryoPDK development. Second, the front end of QNDR1 includes a correlated double-sampling (CDS) circuit to suppress low frequency noise. Because sampling is an inherently nonlinear, time-domain process, traditional AC noise analysis cannot fully capture the noise at the output of the circuit.

\begin{figure}
  \centering
  \includegraphics[width=\linewidth]{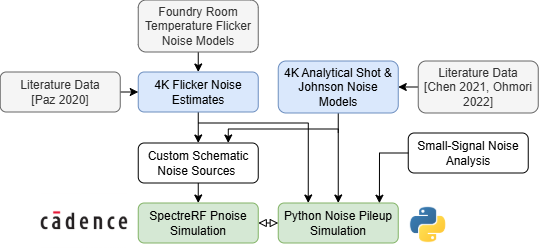}
  \caption{Analysis methodology used to simulate cryogenic noise performance of the QNDR1 preamplifier.}
  \label{fig:noise_analysis}
\end{figure}

We addressed the first difficulty primarily by relying on cryogenic noise data from literature. This approach has inherent limitations as the device dimensions, bias conditions, and test setups reported in literature differ from our own. To attempt to get more relevant data, we have also submitted a test chip, QNDR0 with noise test structures that closely resemble the devices used in QNDR1. We addressed the second difficulty with a two-pronged approach: we simulated the noise of QNDR1 both with SpectreRF Pnoise analysis and with analytical methods in Python. The analytical methods provide only a partial picture of the system noise, but checking that they converge with SpectreRF results gives confidence that the simulator is configured correctly. Our full flow for noise analysis is illustrated in Figure \ref{fig:noise_analysis}. 

Our flow starts by developing literature-informed estimates for device noise contributions. For FET devices operating at cryogenic temperature with moderate drain bias, shot noise rather than thermal noise is the dominant contributor to white noise \cite{Chen_2021} \cite{Ohmori_2023}. We modeled shot noise current density as:

$$S_{I_d}=2qI_d F$$

Where we conservatively assumed a Fano factor of $F=1.0$. 

To estimate flicker noise, we started from the foundry flicker noise models for identical devices operating at room temperature and scaled the noise density by a factor of $10\times$, which approximates results from literature \cite{Paz_2020}. Resistor noise contributions are modeled as Johnson-Nyquist noise based on the resistor's nominal value and the nominal operating temperature.

This process yields a unique estimate of the device noise for every device in the core design. For simulation purposes, these estimates were implemented using ideal noise current sources in parallel with the actual schematic devices. The resulting circuit is simulated with SpectreRF. We used a postprocessing script to discard the erroneous Pnoise contributions from foundry noise models and consider only those from our custom noise models.

In parallel, we performed traditional small signal analysis of the contribution of the primary gain device ($N0$) to the output noise of the preamplifier. A Python script was written to model the effects of sampling on the output noise spectrum. Correlated double sampling with a time delay of $T_d$ can be expressed by the function:

$$\delta (t) - \delta(t-T_d)$$

whose frequency-domain amplitude is given by:

$$H(f) = 2 |\sin (\pi T_d f)|$$

Our Python function convolves this transfer function with the preamplifier's output noise spectrum, obtaining a result which can be compared to the SpectreRF Pnoise result for $N0$.

\begin{figure}
  \centering
  \includegraphics[width=\linewidth]{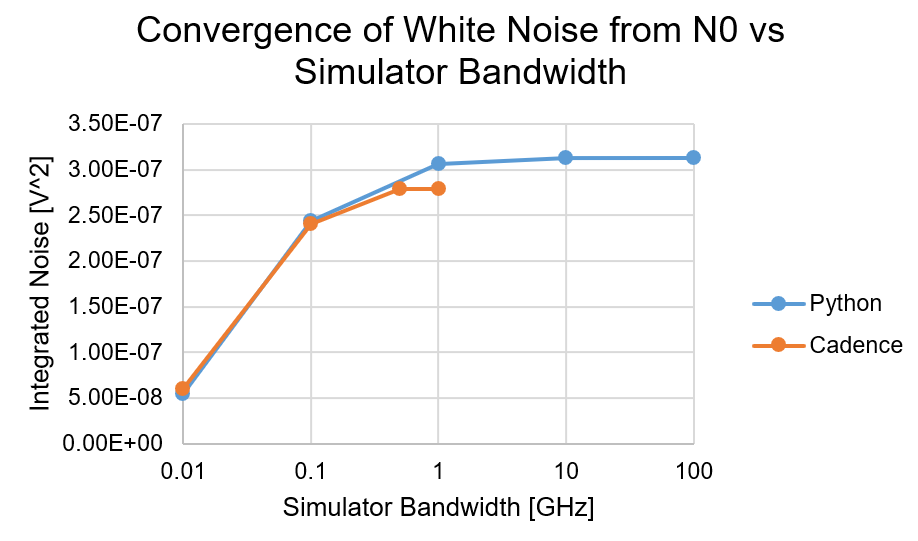}
  \caption{White noise contribution of the preamplifier primary gain device ($N0$) as a function of simulator bandwidth using both the Cadence and Python simulation methods.}
  \label{fig:convergence_correlation}
\end{figure}

To confirm the robustness of our simulation results, we measured the impact of increasing the simulator bandwidth, as shown in Figure \ref{fig:convergence_correlation}. The Cadence SpectreRF simulation can run at a bandwidth of up to $\approx 1 \unit{\giga\hertz}$ before runtime becomes prohibitive. The Python simulation can run at bandwidths $\gt 100\unit{\giga\hertz}$. Given sufficient bandwidth, we observe that the two simulation methods both converge to a fixed result, with a difference of $\approx 10\%$ between the two which we attribute to additional non-dominant poles that are present in the SpectreRF simulation but not the Python simulation.  

\section{Design of QNDR1}

\begin{figure*}
  \centering
  \includegraphics[width=\linewidth]{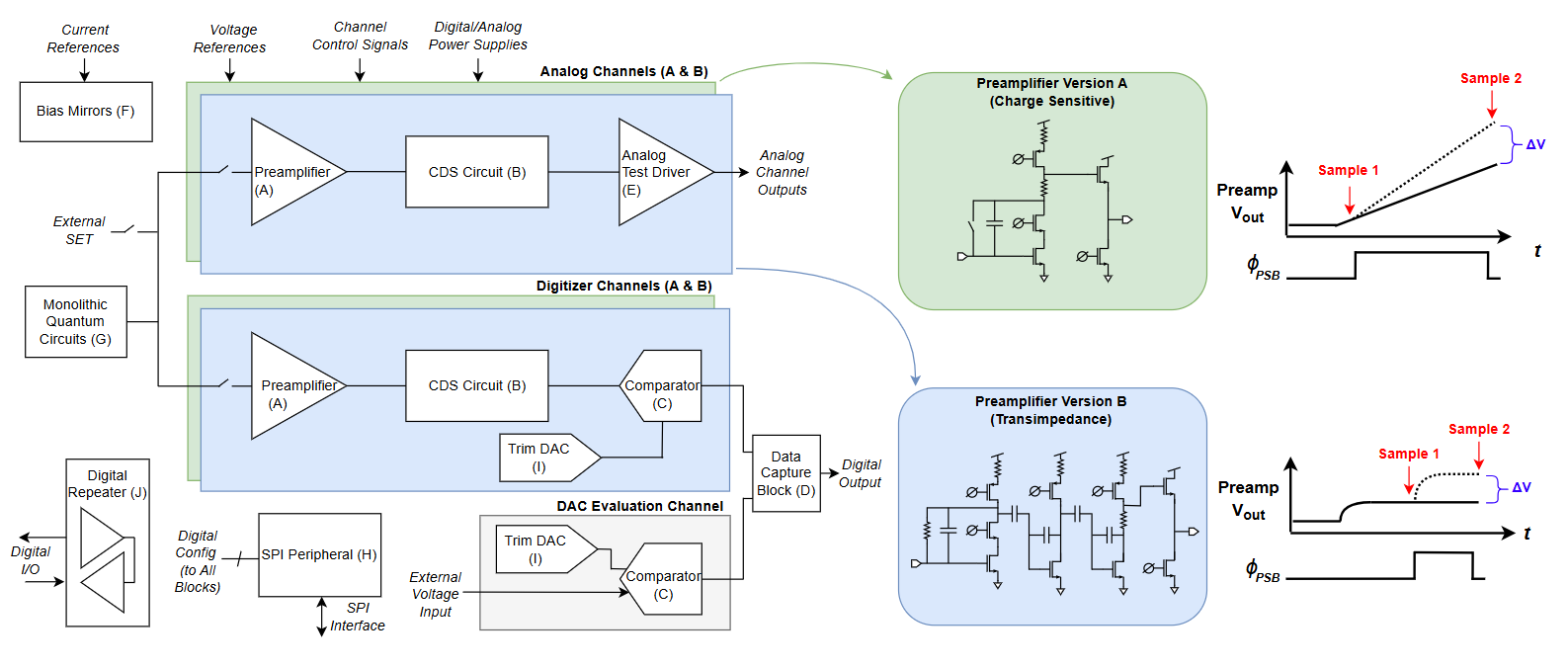}
  \caption{A block diagram of QNDR1 with illustration of the two versions of preamplifier with their respective sampling waveforms.}
  \label{fig:qndr1_bd}
\end{figure*}

QNDR1 is a prototype readout integrated circuit for silicon spin qubits. The primary objective for this chip is to validate the performance of basic circuit blocks which will be the foundation for future highly-scaled readout chips. Because of the high degree of uncertainty in the performance of the quantum dots, single electron transistors, and cryogenic CMOS, QNDR1 is designed with a highly reconfigurable architecture, as shown in Fig. \ref{fig:qndr1_bd}.

The core of QNDR1 consists of four prototype readout channels: two \textbf{Analog Channels} and two \textbf{Digitizer Channels}, any one of which can be connected to the \textbf{Quantum Circuit Block} (\ref{fig:qndr1_bd}(G)) by closing the appropriate switch. The Quantum Circuit Block contains an experimental 22nm FDSOI  double quantum dot coupled to a single electron transistor whose drain is connected to one of the four channels for charge readout experiments. The Quantum Circuit Block can also be bypassed to directly connect the QNDR1 readout channels to a wirebond pad so that they can be used for charge readout of a Quantum Dot + SET on an external chip. 

The following subsections describe the design of the Analog and Digitizer Channels, which are summarized in Table \ref{tab:channel_summary}. Aside from these channels, QNDR1 contains peripheral circuits including on-chip bias mirrors and a SPI peripheral. QNDR1's SPI interface is used to configure the entire chip, including which channel is connected to the Quantum Circuit Block. Non-connected channels may be disabled to conserve power.

\begin{table}
    \centering
    \caption{QNDR1 Channel Variations}
    \begin{tabular}{|c|c|c|}
    \hline
        Channel Num & Preamplifier Version & Channel Type \\
        \hline
        0 & A & Analog\\
        \hline
        1 & B & Analog\\
        \hline
        2 & A & Digitizer\\
        \hline
        3 & B & Digitizer\\
        \hline
    \end{tabular}
    \vspace{0.5em}
    
    \label{tab:channel_summary}
\end{table}

\subsection{QNDR1 Analog Channels}

QNDR1 has two analog channels, which consist of a preamplifier (\ref{fig:qndr1_bd}(A)) followed by a correlated double-sampling (CDS) circuit (\ref{fig:qndr1_bd}(B)) and an analog buffer (\ref{fig:qndr1_bd}(C)).

The preamplifier amplifies the current-mode signal from the SET and converts it to a voltage. The preamplifier also has an important secondary role of providing a low-impedance bias to the drain of the SET, which is described in Section \ref{sec:set_biasing}. The two analog channels of QNDR1 contain two versions of the preamplifier. Preamplifier Version A is a charge-sensitive amplifier , which integrates the SET current onto a feedback capacitor $C_{fb}$. Preamplifier Version B is a resistive feedback transimpedance amplifier followed by two capacitive feedback gain boosting stages. Schematics for both preamplifiers are shown in the right half of Figure \ref{fig:qndr1_bd}.

Both preamplifier versions are followed by the same correlated double sampling circuit, which consists of two sampling capacitors and sampling switches. To read out the spin of an electron in a double quantum dot, a drive voltage $\phi_{PSB}$ is applied to the double quantum dot to induce a spin-dependent charge movement via Pauli Spin Blockade. This causes a spin-dependent change in the SET output current, which appears as a spin-dependent voltage $\Delta V$ at the output of the preamplifier. The CDS circuit samples the output of the preamplifier before and after $\phi_{PSB}$ is applied. Taking the difference between the two samples allows the cancellation of low-frequency and reset noise. The exact timing of the sampling waveforms is dependent on the preamplifier version, as illustrated in Figure \ref{fig:qndr1_bd}.  

After the CDS circuit, analog channels include buffers to drive the sampled voltages directly off-chip for debug and performance characterization.

\subsection{QNDR1 Digitizer Channels}

QNDR has two digitizer channels. The front end of the digitizer channels, like the analog channels, consists of a preamplifier (version A and B) followed by a CDS circuit. Instead of analog buffers, the digitizer channels include a StrongARM-type clocked comparator (\ref{fig:qndr1_bd}(C)) which compares the two CDS samples. A pair of capacitive trim digital-to-analog converters (DACs) (\ref{fig:qndr1_bd}(I)) are used to trim the threshold of the comparator such that it can be used to discriminate between spin up and spin down results. Depending on SPI configuration, the binary output of the comparator can be transmitted directly off-chip, or it can be stored in a set of on-chip registers (\ref{fig:qndr1_bd}(D)) for burst-mode transmission. 

Separate from the four prototype readout channels, QNDR1 also includes a \textbf{DAC Evaluation Channel}, which contains only a comparator and a single trim DAC, allowing independent evaluation of the trim DAC. 

\subsection{SET Biasing in QNDR1}
\label{sec:set_biasing}

In order to achieve the smallest layout area and parasitics, a single electron transistor must be DC-coupled to its readout circuit. This creates a challenge, as the single electron transistor requires a highly accurate DC bias at its drain node to operate in the correct regime, a bias which must be supplied by the readout circuit.

\begin{figure}
  \centering
  \includegraphics[width=\linewidth]{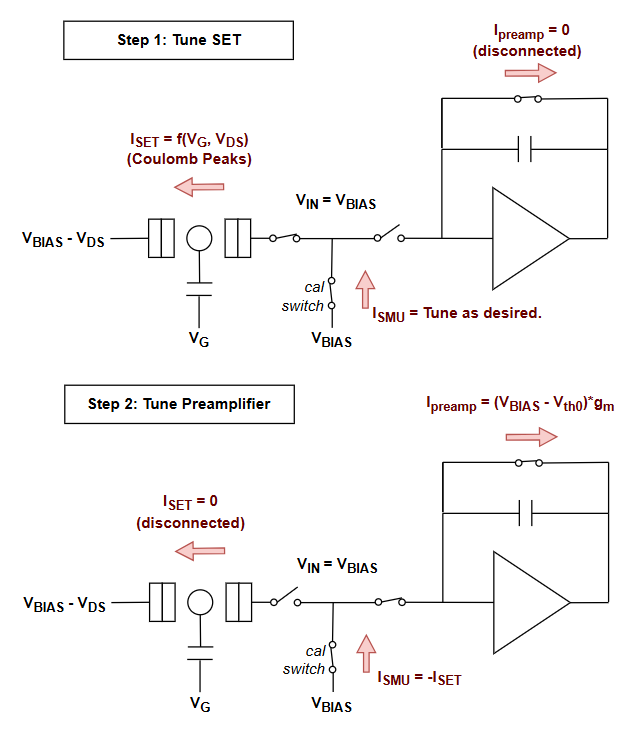}
  \caption{The procedure used in QNDR1 for biasing a single electron transistor DC-coupled to a preamplifier.}
  \label{fig:set_tuneup}
\end{figure}

QNDR1 addresses this challenge using a calibration switch which allows the node in between the SET and the readout circuit to be directly connected to a Source Measure Unit (SMU). The tune-up process is illustrated in Figure \ref{fig:set_tuneup}

First, the SMU is connected to the SET and set to a voltage approximately equal to the threshold voltage of an NMOS transistor. In practice, we selected a voltage of $V_{BIAS}=400\unit{\milli\volt}$ for cryogenic operation. With this node fixed, the source and gate of the SET may be tuned to produce Coulomb peaks. A desired bias point and bias current $I_{SET}$ is found.

Next, the SMU is connected to the input of the preamplifier while retaining the same voltage setting. The preamplifier is put into reset mode, in which case the primary gain device of the preamplifier, $N0$ is effectively diode-connected. The current drawn by the preamplifier from the SMU in this mode is given by:

$$I_{preamp} \approx g_{m0}(V_{BIAS} - V_{th0})$$

Where $V_{th0}$ and $g_{m0}$ are the threshold voltage and transconductance of $N0$. In QNDR1, the back-gate voltage of $N0$, $V_{BGN0}$, is individually controllable, so this knob may be used to tune $V_{th0}$ such that $I_{preamp} = -I_{SET}$.

Once this condition has been achieved, the preamplifier may be directly connected to the SET, and a correct bias will be achieved at an input node voltage of $V_{BIAS}$.

This approach may be scaled to a modest number of channels by supplying $V_{BGN0}$ from an integrated trim DAC per channel. However, scaling to a large array will likely dictate an improvement in SET design to reduce the reliance on fine-tuned bias voltages.

\section{Simulation Results}

The QNDR1 design was simulated using Cadence Spectre. Because of inherent uncertainty about cryogenic models, the design was simulated across three temperature corners: foundry models at 27 C, foundry models at -195 C (the lowest temperature where results converge), and Fermilab's isothermal cryomodels at 3.8K. Together with typical process corners (TT, FF, SF, FS, and SS), this resulted in fifteen simulation corners. To manage the large number of results along with the inherent complexity in the SET biasing procedure described in Section \ref{sec:set_biasing}, we used \textsc{Rhythm}, an open-source Python framework that automates Spectre simulations from the command line \cite{quinn_rhythm}. 

Results for gain, noise and power of the two preamplifier architectures are summarized in Table \ref{tab:sim_summary}.

\begin{table}[!t]
\renewcommand{\arraystretch}{1}
\centering    
\caption{Summary of Simulation Results}
\label{tab:adc_gain}
\begin{tabular}{lcccc}
\hline
\hline
\textsc{Result} &  \textsc{Units} &  \textsc{Spec} & \multicolumn{2}{c}{\textsc{Worst Corner}} \\
 & & &  \textsc{Version A} & \textsc{Version B}\\
\hline
Input Signal &  pA &  450 &  n/a & n/a \\
Input Offset &  pA &  1000 &  n/a & n/a \\
Readout Time &  $\mu s$ &  1.0 &  n/a & n/a \\
ISET Gain &  MV/A &  22.5 &  34.0 & 106.2\\
SNR       & V/V   & 10    & 22.46 & 14.55  \\
Power & $\unit{\micro\watt}$ & 20 & 15.9 & 13.64  \\

\hline
\hline
\end{tabular}
\label{tab:sim_summary}
\end{table}

\begin{figure*}[h]
  \centering
  \includegraphics[width=0.8\linewidth]{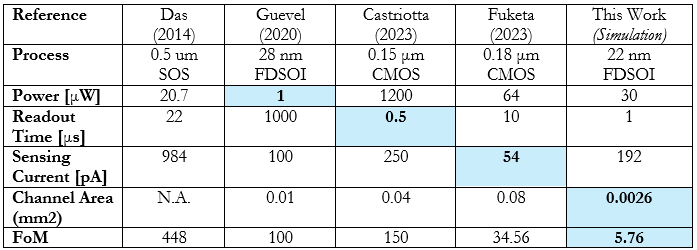}
  \caption{Summary of figures of merit compared to selected recent DC-SET readout integrated circuit designs. References in order of appearance: \cite{Das_2014}\cite{Guevel_2020}\cite{Castriotta_2023}\cite{Fuketa_2023}}
  \label{fig:lit_comparison}
\end{figure*}

\subsection{Gain}

To evaluate the gain of both preamplifier versions, we co-simulated the preamplifier with a traditional SISET using the MIB model. The parameters for this SISET were extracted from Coulomb peak data provided by Diraq based on measurements of actual devices. This data is proprietary and the performance of the resulting SET model differs from the SISET model presented in Section \ref{sec:set_modeling}, which was based on publicly available data from Mahapatra \cite{Mahapatra2004}.

We biased the SET model at a nominal current of $1\unit{\nano\ampere}$ following the procedure in Section \ref{sec:set_biasing} and simulated charge sensing by applying voltage pulses of between $100\unit{\micro\volt}$ and $1\unit{\milli\volt}$ to its gate. The conversion factor between SET gate voltage and drain current is based on uncertain and process-dependent capacitance values, so instead of directly comparing the preamplifier output signal to the SET gate voltage, we measured:

$$\textrm{ISET Gain} = \frac{\textrm{Preamplifier Output Signal [V]}}{\textrm{Average SET }\Delta I_{DS} \textrm{ [A]}}$$

To achieve an output signal of $10 \unit{\milli\volt}$ from a SET signal of $450 \unit{\pico\ampere}$ we required a minimum ISET Gain of $22.5 \frac{MV}{A}$.

\begin{figure}
  \centering
  \includegraphics[width=\linewidth]{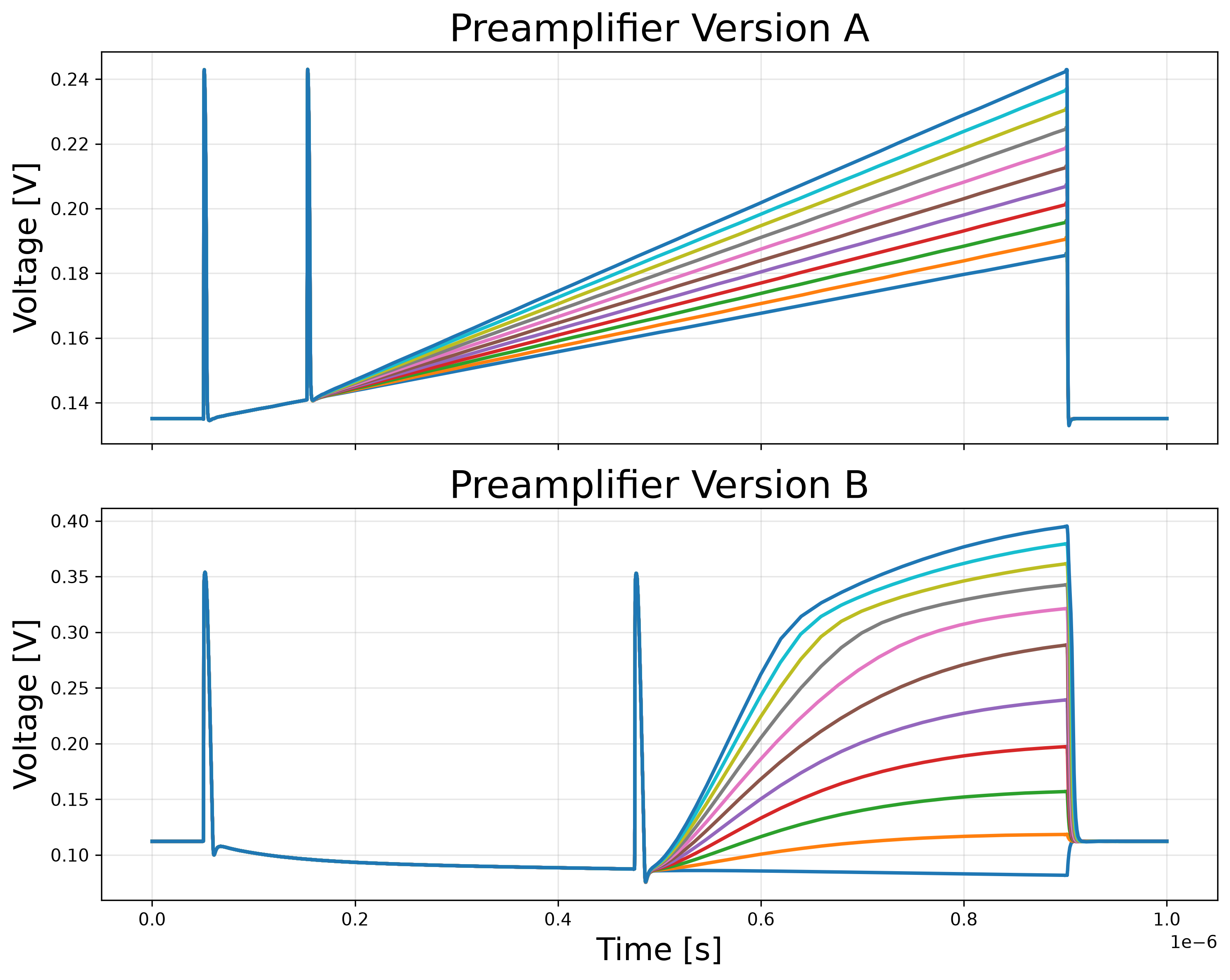}
  \caption{Exemplar simulation traces of $V_{out}$ for Preamplifier Version A and Version B. Traces represent steps of $100\unit{\micro\volt}$ from $0$ to $1\unit{\milli\volt}$.}
  \label{fig:preamp_vout}
\end{figure}

Selected simulation traces are shown in Figure \ref{fig:preamp_vout}

\subsection{Noise}

To evaluate the noise of the QNDR1 front end, we performed SpectreRF periodic steady-state (PSS) and periodic noise (Pnoise) simulations with zero input. Our general approach to achieving valid cryogenic noise simulations is described in Section \ref{sec:cryo_noise}.

Our specification for QNDR1 required a voltage SNR of 10 V/V with respect to a nominal signal of $450 \unit{\pico\ampere}$, which was achieved in all corners. Preamplifier Version B exhibited higher input-referred noise than Version A; in the worst corner, Version B exhibited $32 \unit{\pico\ampere}_{rms}$ input noise, leading to the $\pm 3\sigma $ sensing current of $192 \unit{\pico\ampere}$ reported in Figure \ref{fig:lit_comparison}.

\subsection{Power}

Power is simulated independently per block. Among the core channels, power consumption is dominated by the preamplifier, both versions of which consume $10\sim16\unit{\micro\watt}$ across corners. The comparator, CDS, and other core circuits combined consume $\ll 1 \unit{\micro\watt}$.

The largest power consumption in QNDR1 occurs in the analog test driver and digital output pads due to the need to drive the large capacitive loads presented by cryostat wire harnesses. The analog test driver's bias current can be tuned externally to trade power for bandwidth, and typically falls in the range of $4\sim400\unit{\micro\watt}$ for $\approx 10 \unit{\kilo\hertz} \sim 10\unit{\mega\hertz}$ depending on the simulation corner.

The worst case power consumption of digital output pads is approximated as:

$$C_{load}V_{DDIO}^2f$$

With $V_{DDIO}=1.2V$ and an operational speed of $1 \unit{\mega\hertz}$, the pad can drive $C_{load}=100\unit{\pico\farad}$ with power consumption of $144 \unit{\micro\watt}$, which is within the cooling power of a typical $100\unit{\milli\kelvin}$ stage. In the case of wiring loads $\gt 100\unit{\pico\farad}$, QNDR1 includes a ``digital repeater'' block. This allows a second QNDR1 die to be placed at a higher-power stage of the cryostat and serve to buffer the digital output signal from $\unit{\milli\kelvin}$ to room temperature.

\subsection{Comparison with Literature}

Figure \ref{fig:lit_comparison} compares our simulated results to those achieved in the literature. We follow convention by reporting the minimum readout time and sensing current required for our readout circuit to distinguish between spin-up and spin-down with $3\sigma$ or 99.9\% confidence and by adopting the following Figure of Merit \cite{Fuketa_2023}:

$$\textrm{FoM} = \frac{\textrm{Sensing Current [\unit{\nano\ampere}]}\times\textrm{Power [\unit{\micro\watt}]}}{\textrm{Readout Time [\unit{\micro\second}]}}$$

Although based only on simulation, our results suggest a substantial improvement in both area and overall power-speed-sensitivity compared to the state of the art.

\section{Conclusion}

We have presented Verilog-A models for co-simulating multiple variations of single electron transistors with CMOS readout circuits, including what is to our knowledge the first ever compact model for an asymmetric sensing dot. We have also presented details of our techniques for modeling cryoCMOS circuits, in particular noise analysis at cryogenic temperatures. Finally, we have presented the design of a prototype ASIC for the charge readout of SiMOS quantum dots via a single electron transistor which appears to improve substantially on the state of the art. 

The readout circuits in QNDR1 are intended to serve as a proof of concept, and we believe their area and power consumption leave room for substantial optimization. Future work will include adapting the QNDR1 front end to a low-power pixel format which may be heterogeneously integrated with a quantum IC and may take advantage of the SET variations we have modeled.

QNDR1 is scheduled for tapeout in Fall 2026.

\section*{Acknowledgment}

The authors would like to acknowledge Jonathan Huang, Ajit Dash, and Santiago Serrano Ramirez for enlightening conversations and advice regarding the physics of single electron transistor modeling.

\bibliographystyle{IEEEtran}
\bibliography{IEEEabrv,biblio}
\end{document}